\documentclass[aps,prb,twocolumn,superscriptaddress,floatfix]{revtex4}
\usepackage{times}
\usepackage{graphicx}
\usepackage{epsfig}
\usepackage{amsmath}
\usepackage{amssymb}
\usepackage[usenames, dvipsnames]{color}
\usepackage[version=3]{mhchem}
\usepackage{rotating}

\begin{document}

\title{A 3D-$\Delta$PDF-like measure for hidden anisotropy in disordered materials}

\author{Helena Barker}
\affiliation{Inorganic Chemistry Laboratory, University of Oxford, Oxford, U.K.}

\author{Thomas C. Nicholas}
\affiliation{Center for Molecular Modeling (CMM), Ghent University, Ghent, Belgium}

\author{Andrew L. Goodwin}
\email{andrew.goodwin@chem.ox.ac.uk}
\affiliation{Inorganic Chemistry Laboratory, University of Oxford, Oxford, U.K.}

\date{\today}

\maketitle

For many disordered systems, it is possible to identify structural components with specific local symmetries that are lower than those implied by the symmetry of the bulk phase in which they occur.\cite{Ziman_1979,Simonov_2020} A simple and intuitive example is the case of individual water molecules in liquid water, whereby (on taking time averages) each molecule experiences an environment with $C_{2v}$ point symmetry despite the spherical symmetry of the bulk liquid itself.\cite{Soper_2000} Likewise in the degenerate manifold of pyrochlore spin-ice ground states, where the famous two-in-two-out rules governing spin arrangements within tetrahedral units necessarily break tetrahedral symmetry in a locally uniform manner.\cite{Bramwell_2001} When studying such systems, it may prove interesting to determine over what length-scale and in what manner such local symmetry-breaking meaningfully persists.

In this brief note we present a simple formalism for quantifying such local anisotropy in disordered systems. The approach we take is developed in close analogy to the 3D-$\Delta$PDF, which is a measure of pairwise correlations obtained \emph{via} Fourier transform of the diffuse scattering component in total scattering measurements (\emph{e.g.}\ as performed using X-ray, neutron, or electron diffraction).\cite{Weber_2012} Our purpose here is simply to introduce the formalism, such that it might be used, and we intend to follow up at a later date with a more in-depth study of its application to a variety of canonical disordered systems. Likewise we are conscious that a number of (often closely) related metrics are used in related fields,\cite{Svishchev_1995,Mickel_2013,Bernard_2011,Fijan_2026} and we defer until that later report any proper consideration of the appropriate similarities and differences.

As context we summarise first the key equations that describe the 3D-$\Delta$PDF. A scattering experiment measures the total structure factor $I(\mathbf Q)$ over some range of scattering vectors $\mathbf Q$ determined by the particulars of the experiment. Taking the inverse Fourier transform of this function yields the (three-dimensional) pair distribution function\cite{Schaub_2007}
\begin{equation}
G(\mathbf r)=\mathcal F^{-1}[I(\mathbf Q)].\label{g}
\end{equation}
For crystalline materials, $I(\mathbf Q)$ is separable into Bragg and diffuse scattering contributions
\begin{equation}
I(\mathbf Q)=I_{\rm Bragg}(\mathbf Q)+I_{\rm diff}(\mathbf Q),
\end{equation}
such that the corresponding contributions to $G(\mathbf r)$ might also be determined. The inverse Fourier transform of the Bragg component is of course the Patterson function
\begin{equation}
P(\mathbf r)=\mathcal F^{-1}[I_{\rm Bragg}(\mathbf Q)],
\end{equation}
which is a scattering-weighted pairwise correlation function capturing information about the distribution of interatomic vectors based purely on the average structure (\emph{i.e.}\ the average-structure autocorrelation function). The 3D-$\Delta$PDF is obtained instead from the diffuse scattering as
\begin{equation}
\Delta G(\mathbf r)=\mathcal F^{-1}[I_{\rm diff}(\mathbf Q)]=G(\mathbf r)-P(\mathbf r),\label{dg}
\end{equation}
and quantifies the difference between pairwise vector correlations as they occur locally and as given by the Patterson function.\cite{Weber_2012,Simonov_2014}

Although both $G(\mathbf r)$ and $\Delta G(\mathbf r)$ contain the same information regarding local structure (\emph{i.e.}\ departures from the average), this information is much clearer in the latter than the former because the average correlations no longer dominate the signal.\cite{Weber_2012} We note that these functions can be expanded further in terms of their contributions from different atom-type pairs; for example,
\begin{equation}
\Delta G(\mathbf r)=\sum_{ij}w_{ij}\Delta G_{ij}(\mathbf r).
\end{equation}
Here, $i$ and $j$ denote atom types, and $w_{ij}$ is a weighting that will depend on the corresponding scattering cross sections and relative fractions of these atom-type pairs.\cite{Keen_2001,Kopecky_2004}

Because Fourier transformation and its inverse both preserve point symmetries, each of the functions $G(\mathbf r)$, $P(\mathbf r)$, and $\Delta G(\mathbf r)$ inherits the point symmetry of the scattering function $I(\mathbf Q)$; for crystals this is of course the Laue symmetry. In the case of isotropic (non-crystalline) systems, there is no Bragg scattering nor any orientational dependence; consequently, we have $I(\mathbf Q)=I_{\rm diff}(\mathbf Q)=I(Q)$ and $G(\mathbf r)=\Delta G(\mathbf r)=G(r)$.

\begin{figure}[t]
\begin{center}
\includegraphics{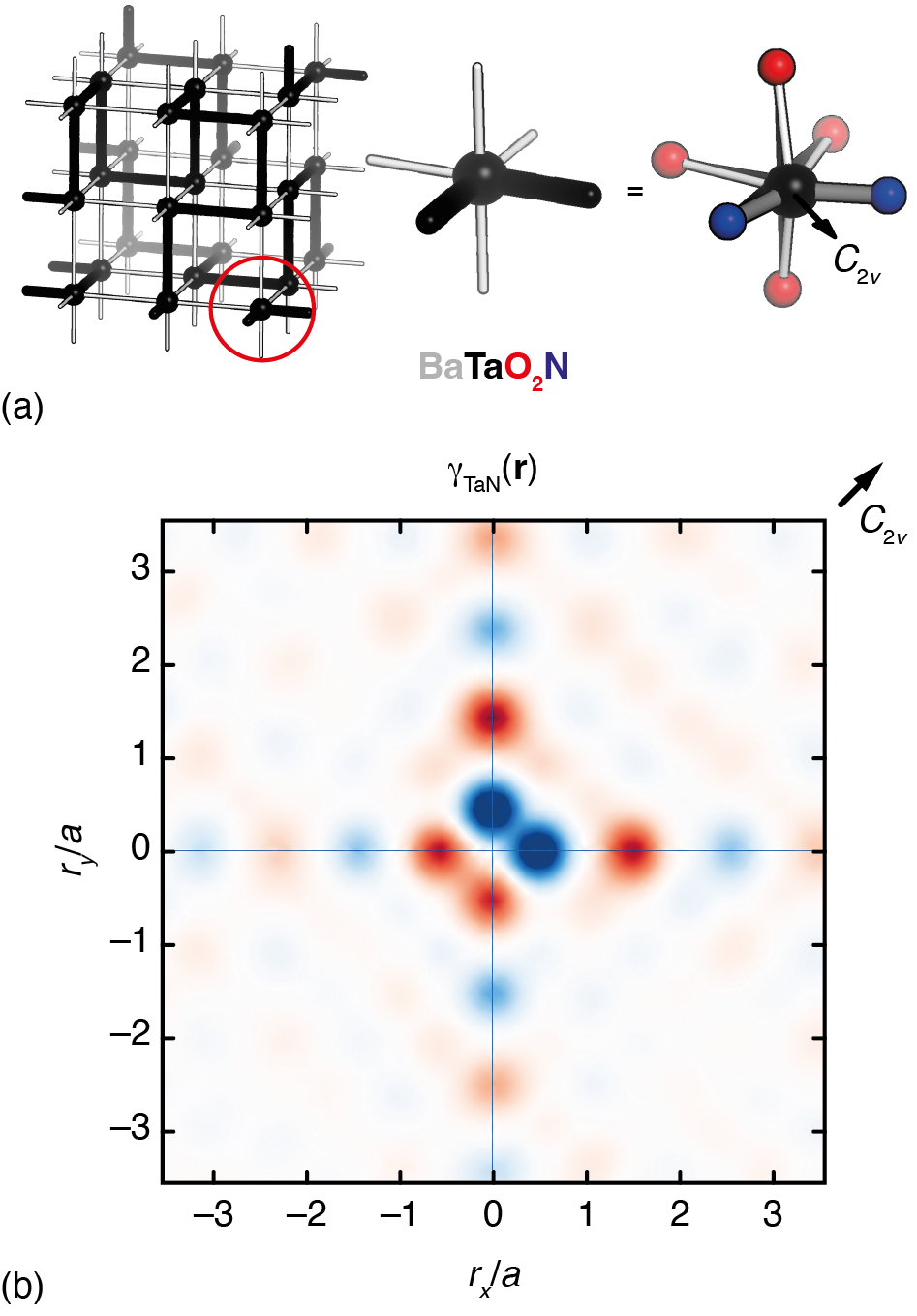}
\end{center}
\caption{\label{fig1} (a) The mixed-anion perovskite BaTaO$_2$N forms as a disordered crystal in which O and N atoms (red and blue spheres, respectively) decorate the same crystallographic site, subject to a strict local rule that each Ta centre (black sphere) is coordinated by exactly two N atoms and exactly four O atoms.\cite{Camp_2012} Ba atoms are omitted from this representation. The local TaO$_4$N$_2$ coordination environment always adopts a \emph{cis} geometry, such that each Ta centre experiences a local environment with $C_{2v}$ point symmetry. The Ta--N--Ta linkages decorate the underlying cubic network topology with a labyrinthine motif (thick black sticks). (b) The local anisotropy function $\gamma_{\rm TaN}(\mathbf r)$ obtained from atomistic representations of this structure shows the spatial distribution of increased (blue) and decreased (red) likelihood of observing N atoms surrounding Ta centres, with reference to a local coordinate system in which the diad axis lies along the local $x=y$ direction. Note (i) that the local anisotropy persists beyond nearest-neighbour pairs and (ii) that the $\gamma_{\rm TaN}(\mathbf r)$ function vanishes on applying the Laue symmetry ($O_h$).}
\end{figure}

We now consider a structural model in which some component (labelled $i$, say) adopts a specific local symmetry that is different from that implied by the average structure. We use the tilde in the notation
\begin{equation}
\tilde G_{ij}(\mathbf r)
\end{equation}
to denote the spatial arrangement of atom types $j$ around atom type $i$ with reference to a suitable local coordinate system that reflects the particular symmetry of interest. In the case of the perovskite-structured BaTaO$_2$N, for example, we might consider the \emph{cis}-arrangement of N atoms around Ta, using a coordinate system in which the Ta--N bonds are assigned as local $x$ and $y$, and the perpendicular axis as local $z$ [Fig.~\ref{fig1}(a)]. This arrangement has $C_{2v}$ point symmetry, a subgroup of the $O_h$ point symmetry of the Ta environment within the global coordinate system. There are $12=|O_h|/|C_{2v}|$ possible arrangements of the local coordinates, corresponding to different orientations of the N--Ta--N triplets in the disordered BaTaO$_2$N structure.

It is possible then to define an equivalent local-coordinate average-structure correlation function
\begin{equation}
\tilde P_{ij}(\mathbf r),
\end{equation}
determined using only information from the Bragg intensities (in the case of an isotropic material this function is of course zero-valued).  And then a local-coordinate 3D-$\Delta$PDF follows naturally:
\begin{equation}
\Delta\tilde G_{ij}(\mathbf r)=\tilde G_{ij}(\mathbf r)-\tilde P_{ij}(\mathbf r).
\end{equation}

We make a few notes at this point. First we observe that whereas the conventional functions \eqref{g}--\eqref{dg} can in principle be measured or derived directly from experiment, the local-coordinate variants depend on access to an atomistic model. Hence the extent to which a $\Delta\tilde G_{ij}(\mathbf r)$ function is physically meaningful will depend entirely on the physicality of the modelling process from which it is obtained. A second point is that $\tilde G_{ij}(\mathbf r)=\Delta\tilde G_{ij}(\mathbf r)$ for isotropic systems. A third is the crucial point that $\Delta \tilde G_{ij}(\mathbf r)$ will contribute to $\Delta G_{ij}(\mathbf r)$---and, hence, to $\Delta G(\mathbf r)$---but contains additional information regarding local anisotropy and its spatial persistence. And a fourth is that application of the Laue symmetry to $\Delta\tilde G_{ij}(\mathbf r)$ recovers $\Delta G_{ij}(\mathbf r)$.

We are now in a position to introduce the local anisotropy measure at the heart of our note. We define
\begin{equation}
\gamma_{ij}(\mathbf r)=\Delta\tilde G_{ij}(\mathbf r)-\Delta G_{ij}(\mathbf r)
\end{equation}
as this anisotropy measure. In words, it is a correlation function that captures the difference in the arrangement of atoms of type $j$ around those of type $i$ when one considers a suitable local coordinate system relative to that imposed by the Laue symmetry. Constructed in this way, $\gamma_{ij}(\mathbf r)$ expressly contains only information about local environments that is not present in the 3D-$\Delta$PDF itself. In particular, it has the property that it vanishes under operation by the elements of the Laue group:
\begin{equation}
\sum_{o\in{\rm L.G.}}o[\gamma_{ij}(\mathbf r)]=0.\label{lg}
\end{equation}
We note the relationship to similar formalisms used in symmetry-adapted strains applied elsewhere.\cite{Thomas_2021} Note that for isotropic systems, $\gamma_{ij}(\mathbf r)$ is no longer itself isotropic but rather inherits the local point symmetry of the site of atom type $i$. Indeed the implication of equation~\eqref{lg} is that
\begin{equation}
\int\gamma_{ij}(\mathbf r)\,{\rm d}\Omega=0,
\end{equation}
where the integral is taken over solid angles $\Omega$. The fact that $\gamma$ characterises local anisotropy but makes no contribution to either the 3D-$\Delta$PDF or the scattering function $I(\mathbf Q)$ is what motivates our use of the term `hidden' for the correlations it contains. A representative example of a $\gamma(\mathbf r)$ function for a model of the BaTaO$_2$N system is shown in Fig.~\ref{fig1}(b).

Because the definition of the local-coordinate frame involves considering the relative positions of groups of more than two atoms, $\gamma$ encodes for higher-order structural correlations. Such correlations are not usually determined uniquely by the two-body correlations measured in a scattering experiment, but may nonetheless be constrained indirectly by them and by additional physical information incorporated in the atomistic model, such as energetics. Accordingly, interpretation of $\gamma$ must take into account both the experimental constraints on the model and any assumptions introduced during its construction.

\acknowledgments
We acknowledge financial support from the UKRI through Frontier Research Grant EP/Z534031/1, from the Royal Society through the Faraday Discovery Fellowships Fund, provided by DSIT, and from the Research Foundation-Flanders (FWO Junior
Postdoctoral Fellowship to T.C.N., grant no. 1264226N).

\bibliography{arxiv_2026_local3d-dpdf}

\begin{thebibliography}{15}
\expandafter\ifx\csname natexlab\endcsname\relax\def\natexlab#1{#1}\fi
\expandafter\ifx\csname bibnamefont\endcsname\relax
  \def\bibnamefont#1{#1}\fi
\expandafter\ifx\csname bibfnamefont\endcsname\relax
  \def\bibfnamefont#1{#1}\fi
\expandafter\ifx\csname citenamefont\endcsname\relax
  \def\citenamefont#1{#1}\fi
\expandafter\ifx\csname url\endcsname\relax
  \def\url#1{\texttt{#1}}\fi
\expandafter\ifx\csname urlprefix\endcsname\relax\def\urlprefix{URL }\fi
\providecommand{\bibinfo}[2]{#2}
\providecommand{\eprint}[2][]{\url{#2}}

\bibitem[{\citenamefont{Ziman}(1979)}]{Ziman_1979}
\bibinfo{author}{\bibfnamefont{J.~M.} \bibnamefont{Ziman}},
  \emph{\bibinfo{title}{Models of disorder. The theoretical physics of
  homogeneously disordered systems}} (\bibinfo{publisher}{Cambridge University
  Press}, \bibinfo{address}{Cambridge}, \bibinfo{year}{1979}).

\bibitem[{\citenamefont{Simonov and Goodwin}(2020)}]{Simonov_2020}
\bibinfo{author}{\bibfnamefont{A.}~\bibnamefont{Simonov}} \bibnamefont{and}
  \bibinfo{author}{\bibfnamefont{A.~L.} \bibnamefont{Goodwin}},
  \bibinfo{journal}{Nat. Rev. Chem.} \textbf{\bibinfo{volume}{4}},
  \bibinfo{pages}{657} (\bibinfo{year}{2020}).

\bibitem[{\citenamefont{Soper and Ricci}(2000)}]{Soper_2000}
\bibinfo{author}{\bibfnamefont{A.~K.} \bibnamefont{Soper}} \bibnamefont{and}
  \bibinfo{author}{\bibfnamefont{M.~A.} \bibnamefont{Ricci}},
  \bibinfo{journal}{Phys. Rev. Lett.} \textbf{\bibinfo{volume}{84}},
  \bibinfo{pages}{2881} (\bibinfo{year}{2000}).

\bibitem[{\citenamefont{Bramwell and Gingras}(2001)}]{Bramwell_2001}
\bibinfo{author}{\bibfnamefont{S.~T.} \bibnamefont{Bramwell}} \bibnamefont{and}
  \bibinfo{author}{\bibfnamefont{M.~J.~P.} \bibnamefont{Gingras}},
  \bibinfo{journal}{Science} \textbf{\bibinfo{volume}{294}},
  \bibinfo{pages}{1495} (\bibinfo{year}{2001}).

\bibitem[{\citenamefont{Weber and Simonov}(2012)}]{Weber_2012}
\bibinfo{author}{\bibfnamefont{T.}~\bibnamefont{Weber}} \bibnamefont{and}
  \bibinfo{author}{\bibfnamefont{A.}~\bibnamefont{Simonov}},
  \bibinfo{journal}{Z. Kristallogr.} \textbf{\bibinfo{volume}{227}},
  \bibinfo{pages}{238} (\bibinfo{year}{2012}).

\bibitem[{\citenamefont{Svishchev and Kusalik}(1995)}]{Svishchev_1995}
\bibinfo{author}{\bibfnamefont{I.~M.} \bibnamefont{Svishchev}}
  \bibnamefont{and} \bibinfo{author}{\bibfnamefont{P.~G.}
  \bibnamefont{Kusalik}}, \bibinfo{journal}{Chem. Phys. Lett.}
  \textbf{\bibinfo{volume}{239}}, \bibinfo{pages}{349} (\bibinfo{year}{1995}).

\bibitem[{\citenamefont{Mickel et~al.}(2013)\citenamefont{Mickel, Kapfer,
  Schroder-Turk, and Mecke}}]{Mickel_2013}
\bibinfo{author}{\bibfnamefont{W.}~\bibnamefont{Mickel}},
  \bibinfo{author}{\bibfnamefont{S.~C.} \bibnamefont{Kapfer}},
  \bibinfo{author}{\bibfnamefont{G.~E.} \bibnamefont{Schroder-Turk}},
  \bibnamefont{and} \bibinfo{author}{\bibfnamefont{K.}~\bibnamefont{Mecke}},
  \bibinfo{journal}{J. Chem. Phys.} \textbf{\bibinfo{volume}{138}},
  \bibinfo{pages}{044501} (\bibinfo{year}{2013}).

\bibitem[{\citenamefont{Bernard and Krauth}(2011)}]{Bernard_2011}
\bibinfo{author}{\bibfnamefont{E.~P.} \bibnamefont{Bernard}} \bibnamefont{and}
  \bibinfo{author}{\bibfnamefont{W.}~\bibnamefont{Krauth}},
  \bibinfo{journal}{Phys. Rev. Lett.} \textbf{\bibinfo{volume}{107}},
  \bibinfo{pages}{155704} (\bibinfo{year}{2011}).

\bibitem[{\citenamefont{Fijan et~al.}(2026)\citenamefont{Fijan, Ward~Rashidi,
  Bradley, and Glotzer}}]{Fijan_2026}
\bibinfo{author}{\bibfnamefont{D.}~\bibnamefont{Fijan}},
  \bibinfo{author}{\bibfnamefont{M.~R.} \bibnamefont{Ward~Rashidi}},
  \bibinfo{author}{\bibfnamefont{J.}~\bibnamefont{Bradley}}, \bibnamefont{and}
  \bibinfo{author}{\bibfnamefont{S.~C.} \bibnamefont{Glotzer}},
  \bibinfo{journal}{J. Chem. Phys.} \textbf{\bibinfo{volume}{164}}
  (\bibinfo{year}{2026}).

\bibitem[{\citenamefont{Schaub et~al.}(2007)\citenamefont{Schaub, Weber, and
  Steurer}}]{Schaub_2007}
\bibinfo{author}{\bibfnamefont{P.}~\bibnamefont{Schaub}},
  \bibinfo{author}{\bibfnamefont{T.}~\bibnamefont{Weber}}, \bibnamefont{and}
  \bibinfo{author}{\bibfnamefont{W.}~\bibnamefont{Steurer}},
  \bibinfo{journal}{Phil. Mag.} \textbf{\bibinfo{volume}{87}},
  \bibinfo{pages}{2781} (\bibinfo{year}{2007}).

\bibitem[{\citenamefont{Simonov et~al.}(2014)\citenamefont{Simonov, Weber, and
  Steurer}}]{Simonov_2014}
\bibinfo{author}{\bibfnamefont{A.}~\bibnamefont{Simonov}},
  \bibinfo{author}{\bibfnamefont{T.}~\bibnamefont{Weber}}, \bibnamefont{and}
  \bibinfo{author}{\bibfnamefont{W.}~\bibnamefont{Steurer}},
  \bibinfo{journal}{J. Appl. Cryst.} \textbf{\bibinfo{volume}{47}},
  \bibinfo{pages}{2011} (\bibinfo{year}{2014}).

\bibitem[{\citenamefont{Keen}(2001)}]{Keen_2001}
\bibinfo{author}{\bibfnamefont{D.~A.} \bibnamefont{Keen}}, \bibinfo{journal}{J.
  Appl. Cryst.} \textbf{\bibinfo{volume}{34}}, \bibinfo{pages}{172}
  (\bibinfo{year}{2001}).

\bibitem[{\citenamefont{Kopeck{\'y}}(2004)}]{Kopecky_2004}
\bibinfo{author}{\bibfnamefont{M.}~\bibnamefont{Kopeck{\'y}}},
  \bibinfo{journal}{J. Appl. Cryst.} \textbf{\bibinfo{volume}{37}},
  \bibinfo{pages}{711} (\bibinfo{year}{2004}).

\bibitem[{\citenamefont{Thomas et~al.}(2021)\citenamefont{Thomas,
  Raju~Natarajan, and Van~der Ven}}]{Thomas_2021}
\bibinfo{author}{\bibfnamefont{J.~C.} \bibnamefont{Thomas}},
  \bibinfo{author}{\bibfnamefont{A.}~\bibnamefont{Raju~Natarajan}},
  \bibnamefont{and} \bibinfo{author}{\bibfnamefont{A.}~\bibnamefont{Van~der
  Ven}}, \bibinfo{journal}{npj Comp. Mater.} \textbf{\bibinfo{volume}{7}},
  \bibinfo{pages}{164} (\bibinfo{year}{2021}).

\bibitem[{\citenamefont{Camp et~al.}(2012)\citenamefont{Camp, Fuertes, and
  Attfield}}]{Camp_2012}
\bibinfo{author}{\bibfnamefont{P.~J.} \bibnamefont{Camp}},
  \bibinfo{author}{\bibfnamefont{A.}~\bibnamefont{Fuertes}}, \bibnamefont{and}
  \bibinfo{author}{\bibfnamefont{J.~P.} \bibnamefont{Attfield}},
  \bibinfo{journal}{J. Am. Chem. Soc.} \textbf{\bibinfo{volume}{134}},
  \bibinfo{pages}{6762} (\bibinfo{year}{2012}).

\end{thebibliography}

\end{document}